\documentclass[10pt,conference]{IEEEtran}

\usepackage{url}
\usepackage{QC,colordvi}
\usepackage{amsmath,amsfonts}
\usepackage{boldmath}

\long\def\magma#1{}
\let\ds\displaystyle

\def\w{\omega}
\def\wbar{\overline{\omega}}

\def\entspricht{\mathrel{\hat{=}}}

\def\ket#1{|#1\rangle}

\def\C{\mathbb C}
\def\F{\mathbb F}

\def\N{\mathbb N}

\def\Cnot{{\rm CNOT}}
\def\Csign{{\rm CSIGN}}
\def\GL{{\rm GL}}

\newtheorem{theorem}{Theorem}
\newtheorem{definition}[theorem]{Definition}
\newtheorem{lemma}[theorem]{Lemma}
\newtheorem{corollary}[theorem]{Corollary}
\newtheorem{algorithm}[theorem]{Algorithm}

\begin{document}

\title{Non-catastrophic Encoders and Encoder Inverses for Quantum Convolutional Codes}

\author{\authorblockN{Markus Grassl}
\authorblockA{Institut f\"ur Algorithmen und Kognitive Systeme\\
Fakult\"at f\"ur Informatik, Universit\"at Karlsruhe (TH)\\
Am Fasanengarten 5, 76128 Karlsruhe, Germany \\
Email: grassl@ira.uka.de}
\and
\authorblockN{Martin R\"otteler}
\authorblockA{NEC Laboratories America, Inc.\\
4 Independence Way, Suite 200\\
Princeton, NJ 08540, U.S.A.\\
Email: mroetteler@nec-labs.com}}
%

\maketitle

\begin{abstract}
  We present an algorithm to construct quantum circuits for encoding
  and inverse encoding of quantum convolutional codes.  We show that
  any quantum convolutional code contains a subcode of finite index
  which has a non-catastrophic encoding circuit.  Our work generalizes
  the conditions for non-catastrophic encoders derived in a paper by
  Ollivier and Tillich (quant-ph/0401134) which are applicable only
  for a restricted class of quantum convolutional codes.  We also show
  that the encoders and their inverses constructed by our method
  naturally can be applied online, i.\,e., qubits can be sent and
  received with constant delay.
\end{abstract}

\section{Introduction}

Similar to the classical case a quantum convolutional code encodes an
incoming stream of quantum information into an outgoing stream.  A
theory of quantum convolutional codes based on infinite stabilizer
matrices has been developed recently, see \cite{OlTi04}. While some
constructions of quantum convolutional codes are known, see
\cite{Chau:98,Chau:99,AlPa04,FoGu05,FGG05,GraRo05,OlTi03,OlTi04}, some
very basic questions about the structure of quantum convolutional
codes and their encoding circuits have not been addressed so far,
respectively have been addressed only in special cases. In this paper
we focus on the question of which quantum convolutional codes have
non-catastrophic encoders, respectively inverse encoders.

Recall, that classically a code encoded by a catastrophic encoder has
the unwanted property that---after code word estimation---a finite
number of error locations can be mapped by the inverse encoder to an
infinite number of error locations. For classical convolutional codes
it is well-known that the non-catastrophicity condition is a property
of the encoder and not of the code itself.  Indeed, every
convolutional code has both catastrophic and non-catastrophic encoders
and therefore the choice of a good encoder is very important.

In this paper we address the analogous question whether any quantum
convolutional stabilizer code has non-catastrophic encoders and
encoder inverses.  Here the condition to be non-catastrophic has been
shown in \cite{OlTi04} to be that it has a constant depth encoder
whose elementary quantum gates can be arranged in form of a ``pearl
necklace'', i.\,e., a regular structure in which blocks are only
allowed to overlap with their neighbors with possibly some blocks
spaced out.  Furthermore, in \cite{OlTi04} some conditions on the code
have been given under which a non-catastrophic encoder exists.
However, these conditions are quite strict and not applicable to an
arbitrary quantum convolutional code.

Using the matrix description of quantum convolutional stabilizer codes
and transformations on this matrix which preserve the symplectic
orthogonality, we show that a normal form can be achieved which
corresponds to a very simple convolutional code.  Reducing the
dimension of this code by only a bounded factor, we obtain an even
simpler code allowing online encoding and decoding. Furthermore, from
the sequence of transformations one can read off a non-catastrophic
encoder for a subcode of the original code whose dimension is
reduced by the same factor.  Asymptotically, the rate of the subcode
and the original code are the same.

\section{Quantum Convolutional Codes}

Quantum convolutional codes are defined as infinite versions of
quantum stabilizer codes. We briefly recall the necessary definitions
and the polynomial formalism to describe quantum convolutional codes
which was introduced in \cite{OlTi04}.
\begin{definition}[Infinite Pauli Group]\label{def:PauliGroup}
Let 
\[
X{=}\left(\begin{array}{cc} 0 & 1 \\ 1 & 0 \end{array} \right), 
Z{=}\left(\begin{array}{rr} 1 & 0 \\ 0 & -1 \end{array} \right), 
Y {=} X Z {=}\left(\begin{array}{rr} 0 & -1 \\ 1 & 0 \end{array} \right)
\]
be the (real version of the) $2 \times 2$ Pauli matrices. Consider an
infinite set of qubits labeled by the nonnegative integers $\N$. Let
$M \in \{X,Y,Z\}$ be a Pauli matrix. We denote by $M_i$ the
semi-infinite tensor product $I_2 \otimes \ldots \otimes I_2 \otimes M
\otimes I_2 \otimes \ldots$, where $M$ operates on qubit $i$ and $I_2$
denotes the identity matrix of size $2\times 2$.  The group generated
by all $X_i$ and $Z_i$ for $i \in \N$ is called the infinite Pauli
group ${\cal P}_\infty$.  For an element $A = A_1 \otimes A_2 \otimes
\ldots \in {\cal P}_\infty$ the positions in which $A_i$ is not equal
to $\pm I_2$ is called the support of $A$.
\end{definition}

In the theory of block stabilizer codes, the elements of the Pauli
group are labeled by tuples of binary vectors.  Similarly, we can
label the elements of the infinite Pauli group by a tuple of binary
sequences, each of which is represented by a formal power series.
Hence we get the correspondence
\begin{eqnarray*}
  (-1)^c X_{\bm{\alpha}} Z_{\bm{\beta}} &:=& (-1)^c\bigotimes_{\ell\ge 0} X^{\alpha_\ell}Z^{\beta_\ell} \\
  &\entspricht&
  \left(\sum_{\ell\ge 0} \alpha_\ell D^\ell,\sum_{\ell\ge 0} \beta_\ell D^\ell\right)
\end{eqnarray*}
where $c \in \F_2$ and $\bm{\alpha} = \sum_{\ell \geq 0} \alpha_\ell
D^\ell$ and $\bm{\beta} = \sum_{\ell \geq 0} \beta_\ell D^\ell$ are
formal power series with coefficients in $\F_2$.  In this
representation, multiplication of elements of ${\cal P}_\infty$
corresponds to addition of the power series.  Furthermore, shifting an
element $A\in {\cal P}_\infty$ one qubit to the right corresponds to
the multiplication of the power series by $D$.  As we also allow to
shift the operators by a bounded number of qubits to the left, we use
Laurent series instead of power series to represent the elements of
${\cal P}_\infty$.  An element $A\in {\cal P}_\infty$ with finite
support corresponds to a tuple of Laurent polynomials. Recall that the
field of Laurent series in the variable $D$ with coefficients in
$\F_2$ is denoted by $\F_2((D))$ and recall further that it contains
the ring $\F_2[D,D^{-1}]$ of Laurent polynomials.

We are interested in shift invariant abelian subgroups of ${\cal
P}_\infty$, more specifically in those subgroups which can be
generated by a finite number of elements and their shifted versions.
The following definition introduces a shorthand notation for
describing such subgroups.

\begin{definition}[Stabilizer Matrix]\label{def:stab_mat}
  Let ${\cal S}$ be an abelian subgroup of ${\cal P}_\infty$ which has
  trivial intersection with the center of ${\cal P}_\infty$.
  Furthermore, let $\{g_1,g_2,\ldots,g_r\}$ where $g_i=(-1)^{c_i}
  X_{\bm{\alpha}_i}Z_{\bm{\beta}_i}$ with $c_i\in\{0,1\}$ and
  $(\bm{\alpha}_i,\bm{\beta}_i)\in\F_2((D))^n\times\F_2((D))^n$ be a
  minimal set of generators for ${\cal S}$. Then a stabilizer matrix
  of the corresponding quantum convolutional (stabilizer) code ${\cal
    C}$ is a generator matrix of the (classical) additive
  convolutional code $C\subseteq \F_2((D))^n\times\F_2((D))^n$
  generated by $(\bm{\alpha}_i,\bm{\beta}_i)$. We will write this
  matrix in the form
\begin{equation}\label{stab_mat}
S(D)=(X(D)|Z(D))=\left(
\begin{array}{c|c}
\bm{\alpha}_1 & \bm{\beta}_1\\
\vdots & \vdots\\
\bm{\alpha}_r & \bm{\beta}_r
\end{array}
\right)\in\F_2((D))^{r\times 2n}.
\end{equation}
\end{definition}

In what follows we are only interested in those stabilizers which have
a finite description. Hence we will consider only such stabilizer
matrices (\ref{stab_mat}) in which all entries are actually rational
functions, i.\,e., elements of $\F_2(D)$. Eventually, we will require
that all entries have finite support and are hence polynomials.

Alternatively to (\ref{stab_mat}) a quantum convolutional code can
also be described in terms of a semi-infinite stabilizer matrix $S$
which has entries in $\F_2 \times \F_2$.  The general structure of the
matrix is as follows:
\begin{equation}\label{semi-inf}
S:=\left(
\begin{array}{cccccccc}
G_0 & G_1 & \ldots & G_m    & 0      & \ldots &        & \\
0 &   G_0 & G_1    & \ldots & G_m    & 0      & \ldots & \\
0 &   0   & G_0    & G_1    & \ldots & G_m    & 0      & \ldots \\
\vdots &  &        & \ddots & \ddots &        & \ddots &
\end{array}
\right)
\end{equation}
The matrix $S$ has a block band structure where each block is of size
$(n-k)\times (m+1) n$.  All blocks have equal size and are comprised
of $m+1$ matrices $G_0, G_1, \ldots, G_m$ which are of size $(n-k)
\times n$ each.  In the second block, these $m+1$ matrices are shifted
by $n$ columns, hence any two consecutive blocks overlap in $(m-1)n$
positions.

Similar to the classical case the link between the polynomial
description of eq.~(\ref{stab_mat}) and the semi-infinite matrix
eq.~(\ref{semi-inf}) is given by $S(D) := \sum_{i=0}^m G_i D^i$.  The
band structure of eq.~(\ref{semi-inf}) implies that for every qubit in the
semi-infinite stream of qubits, there is a bounded number of
generators of the stabilizer group that act non-trivially on that
position.  Moreover, as these generators of the stabilizer group have
bounded support, their eigenvalues can be measured when the
corresponding qubits have been received.  Therefore, it is possible to
compute the error syndrome for the quantum convolutional code online.

Writing the stabilizer in the form $S(D) = (X(D) | Z(D))$ as in
eq.~(\ref{stab_mat}), it was shown in \cite{OlTi04} that the condition
of symplectic orthogonality of the semi-infinite matrix $S$ can be
expressed compactly in the form
\begin{equation}\label{eq:symp_condition}
X(D) Z(1/D)^t + Z(D) X(1/D)^t = 0.
\end{equation}

On the other hand, we can start with an arbitrary self-orthogonal
additive convolutional code over $\F_2((D))^n\times\F_2((D))^n$ to
define a convolutional quantum code.  In general, the generator matrix
for such a code may contain rational functions, but there is always an
equivalent description in terms of a matrix with polynomial entries
\cite{JoZi99}.  The following theorem shows that for self-dual
convolutional codes, all entries of a systematic generator matrix are
in fact Laurent polynomials.
\begin{theorem}\label{satz:selfdual}
Let $S(D)=(X(D)|Z(D))$ with $X(D)=I$ be a stabilizer matrix of a
self-dual additive convolutional code over the rational function field
$\F_{2}(D)$.  Then $Z(1/D)=Z(D)^t$ and all entries of $Z(D)$ are
Laurent polynomials.
\end{theorem}
\begin{proof}
  From condition (\ref{eq:symp_condition}) it follows that the code is
  self-dual if and only if $Z(1/D)=Z(D)^t$. Assume that $Z_{ij}(D)$ is
  a proper rational function and not a Laurent polynomial. 
  Then evaluating the series expansion of $Z_{ij}(D)$ at $1/D$ yields
  infinitely many negative powers. However, since $Z_{ji}(D)$
  contains only finitely many negative powers we get a
  contradiction. Hence all entries of $Z(D)$ have to be Laurent
  polynomials.

  The symmetry $Z(1/D)=Z(D)^t$ additionally implies that the diagonal
  terms $Z_{ii}(D)$ are Laurent polynomials of the form
\begin{equation}\label{eq:symm_diagonal}
Z_{ii}(D)=\sum_{\ell=0}^d c_\ell(D^{-\ell}+D^\ell).
\end{equation}
\end{proof}

\section{Shift-invariant Clifford Operations}
We are interested in quantum circuits which encode a convolutional
quantum code. Recall that the controlled-not (CNOT) maps
$\ket{x}\ket{y} \mapsto \ket{x}\ket{x \oplus y}$ and that the
controlled-Z (CSIGN) gates maps $\ket{x}\ket{y} \mapsto (-1)^{x \cdot
  y} \ket{x}\ket{y}$ (see \cite{NC:2000}). We want that errors which
happen during the encoding do not be spread out too far. A
particularly bad example of spreading errors is given by the cascade
$\Cnot_\infty = \prod_{i=0}^\infty \Cnot^{(i,i+1)}$ where gates with
smaller index $i$ are applied first.  The cascade $\Cnot_\infty$ maps
the finite support element $X \otimes I_2 \otimes I_2 \otimes \ldots$
to the infinite support element $X \otimes X \otimes X \otimes\ldots$
On the other hand the infinite cascade $\Csign_\infty =
\prod_{i=0}^\infty \Csign^{(i,i+1)}$ does not have this behavior:
indeed, a Pauli matrix $X_i$ is mapped to $Z_{i-1} X_i Z_{i+1}$ by
this cascade and since it furthermore commutes with all $Z$ operators,
this shows that it maps finite support Pauli matrices to finite
support Pauli matrices.  The reason for this difference is that the
sequence $\Csign_\infty$ can be parallelized to have finite depth
(actually depth $2$), whereas this is not possible for $\Cnot_\infty$.
Clearly, any circuit of constant depth only leads to a local error
expansion, i.\,e., Pauli matrices with finite support get mapped onto
Pauli matrices with finite support. This gives rise to the following
definition:
\begin{definition}[Non-catastrophic encoder]
  Let ${\cal C}$ be a quantum convolutional code and let ${\cal E}$ be
  an encoding circuit for ${\cal C}$. Then ${\cal E}$ is called {\em
  non-catastrophic} if the gates in ${\cal E}$ can be arranged into a
  circuit of finite depth.
\end{definition}

In the following, we consider infinite cascades of gates from the
Clifford group that can be realized by quantum circuits with constant
depth. Since the generators for the quantum convolutional code are
obtained by shifting a fixed block an infinite number of times, we
have to impose a shift invariance condition on any Clifford gate that
we intend to apply to the code. This means that whenever a gate is
applied it has to be applied also in a shifted version by an offset of
$n$ qubits. Similar to the approach in \cite{GRB:2003}, the action of
such operations on elements of the infinite Pauli group can be
described as linear transformations on the stabilizer matrix. As an
example, the action of an infinitely replicated Hadamard gate $H$ on a
qubit is described in its action on the vectors $(f(D), g(D)) \in
\F_2(D)^2$
by the matrix $\overline{H} = \left( \begin{array}{cc} 0 & 1 \\
    1 & 0 \end{array} \right)$ since $H^\dagger X H = Z$ and
$H^\dagger Z H = X$. Similarly, all infinitely replicated versions of
Clifford gates which only operate within a block and do not connect
qubits between shifted blocks, correspond to the usual matrices in the
symplectic group ${\rm Sp}_{2n}(\F_2)$.

More interesting are those operations which connect different blocks
which have been shifted in time. An example is a CNOT gate which
operates on a qubit $i$ (control) and qubit $j$ (target), where qubit
$j$ has been shifted by $\ell$ blocks.  Recall that shifting by $\ell$
blocks corresponds to multiplying by $D^\ell$.  In this case we obtain
that CNOT gate maps the stabilizer vector $(x_1,x_2|z_1,z_2) \mapsto
(x_1, x_2+x_1 D^\ell|z_1 + z_2 D^{-\ell}, z_2)$, i.\,e., $X$ errors
are propagated into the future and $Z$ errors into the past. Note that
by applying a sequence of CNOT gates we can actually map
$(x_1,x_2|z_1,z_2) \mapsto (x_1,x_2+ f(D) x_1|z_1 + f(1/D) z_2, z_2)$,
where $f(D) \in \F_2[D]$ is an arbitrary polynomial. A summary
of the gates used is shown in Table~\ref{table:clifford}. It is
important to note that all the operations shown in
Table~\ref{table:clifford} can be parallelized to have constant depth.

\begin{table}
\caption{Action of various Clifford operations. \label{table:clifford}}
\vskip-5ex
\[\def\arraystretch{1.5}
\begin{array}{cc}
\hline
\text{unitary gate $U$} & \text{matrix $\overline{U}$}\\
\hline\\[-2ex]
H=\ds\frac{1}{\sqrt{2}}
  \begin{pmatrix}1&1\\1&-1\end{pmatrix}\in\C^{2\times 2}&
\overline{H}
=\begin{pmatrix}0&1\\1&0\end{pmatrix}\in\F_2^{2\times 2}\\[4ex]
\hline\\[-2ex]
P=\begin{pmatrix}1&0\\0&\exp(i \pi/2)\end{pmatrix}\in\C^{2\times 2}&
\overline{P}
=\begin{pmatrix}1&1\\0&1\end{pmatrix}\in\F_2^{2\times 2}\\[4ex]
\hline\\[-2ex]
\Cnot^{(i,j+\ell n)}, i\not\equiv j\pmod n &
\overline{\Cnot}=
\left(\begin{array}{@{}cc|cc@{}}
1 & D^\ell & 0         & 0\\
0 & 1      & 0         & 0\\
\hline
0 & 0      & 1         & 0\\
0 & 0      & D^{-\ell} & 1
\end{array}\right)\\[7ex]
\hline\\[-2ex]
\text{\Csign}^{(i,j+\ell n)}, i\not\equiv j\pmod n&
\overline{\text{\Csign}}=
\left(\begin{array}{@{}cc|cc@{}}
1 & 0 & 0         & D^\ell\\
0 & 1 & D^{-\ell} & 0     \\
\hline
0 & 0 & 1         & 0     \\
0 & 0 & 0         & 1
\end{array}\right)\\[7ex]
\hline\\[-2ex]
P_\ell:=\text{\Csign}^{(i,i+\ell n)},\ell\ne 0 &
\overline{P_{\ell}}=
\begin{pmatrix}
1 & D^{-\ell}+D^\ell\\
0 & 1
\end{pmatrix}\\[4ex]
\hline
\end{array}
\]
Conjugation of the stabilizer group ${\cal S}$ by the unitary gate $U$
corresponds to the action of the matrices $\overline{U}$ on the
columns of the stabilizer matrix $S(D)=(X(D)|Z(D))$.
\vskip-3ex
\end{table}

\section{Computing an Encoding Circuit}
In the following we describe an algorithm which operate on the
stabilizer matrix (\ref{stab_mat}) in order to produce a new
stabilizer which is in a simpler form. We can act in two ways: (i) by
applying row operations using an invertible matrix over
$\F_2[D,D^{-1}]$. Apart from possible shifts, this does not change the
stabilizer group, i.\,e., up to a possibly new initial qubit sequence
(of bounded length) the quantum code is unchanged.  We can also apply
(ii) column operations given by an arbitrary element of the Clifford
group shown in Table~\ref{table:clifford}. Before we state the
algorithm we recall the Smith normal form \cite{JoZi99} of a matrix:
\begin{theorem} Let $M(D)\in \F_2[D]^{r \times n}$ be an $r \times n$
  polynomial matrix. Then there exist polynomial matrices
  $A(D)\in \GL_r(\F_2[D])$ and $B(D)\in \GL_n(\F_2[D])$, both having
  determinant one, such that $M(D) = A(D) \Gamma(D) B(D)$, where
  $\Gamma(D)$ is the $r\times n$ matrix 
\[
\Gamma(D) = \left(
\begin{array}{cccccc}
\gamma_1(D) & & & & & \\
& \ddots & & & & \\
& & \gamma_r(D) & 0 & \cdots & 0 
\end{array}
\right),
\]
where the diagonal elements (\emph{elementary divisors}) $\gamma_i \in
\F_2[D]$ satisfy $\gamma_i | \gamma_{i+1}$ for $i=1, \ldots, r-1$.
\end{theorem}
Note that the Smith form can be computed for any matrix over an
Euclidean domain, including the ring $\F_2[D,D^{-1}]$ of Laurent
polynomials (see, e.g., \cite{DS98}).  For this, define the
degree of a Laurent polynomial $f=\sum_{\ell=\ell_0}^{\ell_1} c_\ell
D^\ell$ with $c_{\ell_0}\ne 0\ne c_{\ell_1}$ as $|\ell_1-\ell_0|$.

We will also need an observation about matrices which have already been
partially brought into Smith form and which contain Laurent
polynomials as entries.  
\begin{lemma}\label{deg_reduction}
  Let $M(D)\in \F_2[D,D^{-1}]^{r \times n}$ be a matrix containing
  Laurent polynomials and which has the form $M = ({\rm
  diag}(\gamma_i(D)) | U(D))$, where $U(D) \in \F_2[D,D^{-1}]^{r
  \times (n-r)}$. Assume that for at least one $i$ we have that
  $\gamma_i$ does not divide the Laurent polynomials contained in the
  $i$th row of $U(D)$. Then at least one of the polynomials
  $\gamma_i^\prime(D)$ arising in the Smith normal form of $M(D)$
  (after the denominators have been cleared by row-wise multiplication
  of powers of $D$) has a strictly smaller degree than the
  corresponding $\gamma_i(D)$.
\end{lemma}
\begin{proof} Without loss of generality, we consider  the first row
  $(\gamma_1(D), 0, \ldots, 0, f_1(D), \ldots, f_{n-r}(D))$ of $M(D)$,
  where the $f_i(D)$ are Laurent polynomials. Clearing the
  denominators by a suitable power $D^\ell$ leaves us with $D^\ell
  \gamma_1(D)$ and polynomials $f_i^\prime(D) := D^\ell f_i(D)$.
  Computing the Smith normal form we obtain the gcd of
  $D^\ell\gamma_1(D), f_1^\prime(D), \ldots, f_{n-r}^\prime(D)$ which
  by assumption has to be a proper divisor of $\gamma_1(D)$.
\end{proof}

Next, observe that by using Clifford gates (acting on the $X$-part
only) we can implement the matrix $B(D)$ used in the Smith normal
form. The reason for this is that in the computation of the Smith
normal form only elementary operations and permutations are necessary
\cite[Section 2.2]{JoZi99}. We can realize these operations using the
$\overline{\Cnot}$ gates and permutations of the qubits, which can
also be realized by $\overline{\Cnot}$.  Left multiplication by an
invertible matrix does not change the stabilizer, so
 there is no need to implement the matrix $A(D)$ as quantum gates.
\begin{algorithm} 
  Let a polynomial stabilizer matrix $S(D)=(X(D)|Z(D))\in\F_2[D]^{r
    \times 2n}$ of full rank be given.
\begin{enumerate}
\item Compute matrices $A(D)$ and $B(D)$ which realize the Smith
  normal form for $X(D)$. Factor the matrix $B(D)$ into elementary
  matrices of the form $\overline{\Cnot}$ and permutations of qubits.
  Apply these operations to the code to obtain the new stabilizer
  matrix
\[
S(D)=\left(
\begin{array}{c|c}
\begin{array}{@{}cc}
\Gamma(D)&0\\
0&0
\end{array}&
\begin{array}{cc@{}}
Z_1(D)& Z_2(D)
\end{array}
\end{array}
\right),
\]
where $\Gamma(D)$ is a diagonal matrix with non-zero polynomial
entries of rank $s$ and
$Z_1(D)\in\F_2[D,D^{-1}]^{r \times s}$ and $Z_2(D)\in\F_2[D,D^{-1}]^{
  r \times (n-s)}$ are matrices with Laurent polynomials as entries.
\item While the $Z_2(D)$ part of $S(D)$ is not zero, repeat the
  following steps: 
\begin{itemize}
\item Use Hadamard gates $\overline{H}$ to swap $Z_2(D)$ into the
  $X$-part yielding 
\[
S(D)=\left(
\begin{array}{cc|cc}
\begin{array}{@{}c}
\Gamma(D)\\
0
\end{array}&
X_2(D)& Z_1(D)&0
\end{array}
\right),
\]
  with $X_2(D)=Z_2(D)$.
\item If $\Gamma(D)$ has full rank and if all polynomials in row $j$
  of $X_2(D)$ are divisible by $\gamma_j(D)$ for all $j=1, \ldots, r$,
  then use $\overline{\Cnot}$-gates to obtain zeros in both $X_2(D)$
  and $Z_2(D)$.
\item Else recompute the Smith normal form of the $X$-part and get
  either smaller elementary divisors or all polynomials in $Z_2(D)$
  are multiples of the corresponding elementary divisor. The degree of
  the elementary divisors decreases because of Lemma
  \ref{deg_reduction}.
\end{itemize}
\item The stabilizer matrix $S(D)$ is now of the form
  $S(D)=(\Gamma(D)0|Z_1(D) 0)$, where $\Gamma(D)$ has a rational
  inverse since $S(D)$ has full rank.
\item From Theorem~\ref{satz:selfdual} it follows that all
  entries in the rows of $Z_1(D)$ are divisible (as Laurent
  polynomials) by the corresponding element of $\Gamma(D)$ (consider
  the matrix $\Gamma^{-1}S(D)=(I\;0|\Gamma^{-1}Z_1(D)\; 0)$ which
  contains Laurent polynomials only). Hence, using $\overline{\Csign}$
  gates, clear all off-diagonal terms in $Z_1(D)$.
\item From (\ref{eq:symm_diagonal}) it follows that we can cancel the
  diagonal of the matrix $Z_1(D)$ using the gates $\overline{P}$ and
  $\overline{P_\ell}$.
\item Finally, use Hadamard gates $\overline{H}$ to obtain $Z$-only
  generators in diagonal form.
\end{enumerate}
\end{algorithm}
This algorithm transforms the original stabilizer matrix into a
stabilizer matrix $S_1(D):=(0\,0|\Gamma(D)0)$ with $\Gamma(D)={\rm
  diag}(\gamma_i(D))$. In case $\gamma_i(D)=1$, the only possible
sequence of states formed by the $i^\text{th}$ qubit of all blocks is
$\ket{0}\ket{0}\ldots$ If $\gamma_i(D)=D^\ell$, there are no
constraints on the first $\ell$ qubits. Otherwise, the state
$\ket{c_0}\ket{c_1}\ldots$ corresponding to the power series expansion
of $1/\gamma_i(D)=\sum_{\ell\ge 0}c_\ell D^\ell$ and its shifted
versions are allowed, too.  As the sequence $(c_\ell)_\ell$ is
periodic, there are only finitely many different shifted versions.  We
ignore these additional states as they would require an infinite
cascade of $\Cnot$ gates. As an example for this behavior consider the
states $\ket{0}\ket{0}\ldots$ and $\ket{1}\ket{1}\ldots$ allowed by
the single qubit $Z$-generator $1{+}D$.  

In case $\Gamma(D)\ne I$, which corresponds to catastrophic encoders
in the classical case, we consider the code ${\cal C}_0$ with
stabilizer matrix $S_0(D):=(0\,0|I\,0)$. Now, ${\cal C}_0$ is a proper
convolutional subcode of the code ${\cal C}_1$ with stabilizer matrix
$S_1(D)$.  The dimension is only decreased by a bounded factor
depending on $\Gamma(D)$. In case $\Gamma(D)=I$, we have ${\cal
  C}_0={\cal C}_1$.

The subcode ${\cal C}_0$ has a very simple structure: a sequence of
$n-k$ qubits in the state $\ket{0}$ alternates with a sequence of $k$
qubits $\ket{\phi_i}$.  Encoding for ${\cal C}_0$ is done by inserting
qubits in the state $\ket{0}$ into the input stream.  To obtain a
state of (a subcode of) the original convolutional quantum code ${\cal
  C}$, apply the gates corresponding to the elementary matrices used
in the algorithm in reversed order.  The corresponding elementary
gates are only Clifford gates which have to be replicated infinitely
often.  All elementary gates used can be parallelized into finite
depth which implies that the operations can be carried out online.
Hence we have shown the following result:

\begin{figure*}[t]
\centerline{\unitlength0.8\unitlength
\inputwires[$\ket{0}$,$\ket{0}$,$\ket{0}$,
$\ket{0}$,$\ket{0}$,$\ket{\phi_1}$,
$\ket{0}$,$\ket{0}$,$\ket{\phi_2}$,
$\ket{0}$,$\ket{0}$,$\ket{\phi_3}$,$\;\vdots\;$](13)
\rlap{\OneQubitGate[12](1,1){$H$}}%
\rlap{\OneQubitGate[10](1,2){$H$}}%
\rlap{\OneQubitGate[9](1,1){$H$}}%
\rlap{\OneQubitGate[7](1,2){$H$}}%
\rlap{\OneQubitGate[6](1,1){$H$}}%
\rlap{\OneQubitGate[4](1,2){$H$}}%
\rlap{\OneQubitGate[3](1,1){$H$}}%
\rlap{\OneQubitGate[1](1,2){$H$}}%
\OneQubitGate(1,1){$H$}
\rlap{\OneQubitGate[10](1,3){$P$}}%
\rlap{\OneQubitGate[7](1,3){$P$}}%
\rlap{\OneQubitGate[4](1,3){$P$}}%
\rlap{\OneQubitGate[1](1,3){$P$}}%
\OneQubitGate(1,1){$P$}
\wires[10](13)
\rlap{\CGate(2,9,4){$Z$}}%
\rlap{\raisebox{20\unitlength}{\CGate(2,3,3){$Z$}}}%
\rlap{\raisebox{60\unitlength}{\CGate(2,9,10){$Z$}}}%
\rlap{\raisebox{140\unitlength}{\CGate(2,6,6){$Z$}}}%
\raisebox{200\unitlength}{\CGate(2,3,3){$Z$}}%
\rlap{\CGate(5,12,13){$Z$}}%
\rlap{\raisebox{80\unitlength}{\CGate(2,6,6){$Z$}}}%
\raisebox{140\unitlength}{\CGate(2,3,3){$Z$}}%
\rlap{\CGate(8,15,13){$Z$}}%
\rlap{\raisebox{20\unitlength}{\CGate(2,6,6){$Z$}}}%
\raisebox{80\unitlength}{\CGate(2,3,3){$Z$}}%
\wires[10](13)
\rlap{\CGate(1,3,1){$Z$}}%
\rlap{\CGate(3,1,3){$Z$}}%
\rlap{\CGate(2,1,2){$Z$}}%
\rlap{\raisebox{60\unitlength}{\CGate(4,9,10){$Z$}}}%
\rlap{\raisebox{140\unitlength}{\CGate(4,6,6){$Z$}}}%
\rlap{\raisebox{180\unitlength}{\CGate(4,2,4){$Z$}}}%
\raisebox{180\unitlength}{\CGate(2,1,2){$Z$}}%
\rlap{\CGate(7,12,13){$Z$}}%
\rlap{\raisebox{80\unitlength}{\CGate(4,6,6){$Z$}}}%
\rlap{\raisebox{120\unitlength}{\CGate(4,2,4){$Z$}}}%
\raisebox{120\unitlength}{\CGate(2,1,2){$Z$}}%
\rlap{\CGate(10,15,13){$Z$}}%
\rlap{\raisebox{20\unitlength}{\CGate(4,6,6){$Z$}}}%
\rlap{\raisebox{60\unitlength}{\CGate(4,2,4){$Z$}}}%
\raisebox{60\unitlength}{\CGate(2,1,2){$Z$}}%
\wires[10](13)
\rlap{\CNOT(2,3,13)}%
\rlap{\CNOT(2,3,10)}%
\rlap{\CNOT(2,3,7)}%
\CNOT(2,3,4)\kern-5\unitlength
\rlap{\CNOT(3,5,13)}%
\rlap{\CNOT(3,5,10)}%
\rlap{\CNOT(3,5,7)}%
\CNOT(3,5,4)\kern-5\unitlength
\rlap{\CNOT(2,6,13)}%
\CNOT(2,6,7)\kern-15\unitlength
\rlap{\CNOT(5,9,13)}%
\CNOT(5,9,7)
\wires[10](13)
\rlap{\CNOT(1,2,13)}%
\rlap{\CNOT(1,2,10)}%
\rlap{\CNOT(1,2,7)}%
\rlap{\CNOT(1,2,4)}%
\CNOT(1,2,1)\kern-5\unitlength
\rlap{\CNOT(1,3,13)}%
\rlap{\CNOT(1,3,10)}%
\rlap{\CNOT(1,3,7)}%
\rlap{\CNOT(1,3,4)}%
\CNOT(1,3,1)\kern-5\unitlength
\rlap{\CNOT(1,6,13)}%
\rlap{\CNOT(1,6,7)}%
\CNOT(1,6,1)\kern-15\unitlength
\rlap{\CNOT(4,9,13)}%
\CNOT(1,6,4)\kern-5\unitlength
\rlap{\CNOT(2,4,13)}%
\rlap{\CNOT(2,4,10)}%
\rlap{\CNOT(2,4,7)}%
\CNOT(2,4,4)
%
%
\begin{picture}(0,215)
\put(-120,-35){\dashbox(120,300){}}
\put(-220,-35){\dashbox(90,300){}}
\put(-325,-35){\dashbox(90,300){}}
\put(-425,-35){\dashbox(90,300){}}
\put(-495,-35){\dashbox(60,300){}}
\end{picture}%
%
%
\begin{picture}(0,0)
\put(5,-5){\makebox(0,0)[rt]{\White{\rule{510\unitlength}{100\unitlength}}}}
\end{picture}%
\outputwires[,$\left.\rule{0pt}{30\unitlength}\right\}$ \small block 1,,
,$\left.\rule{0pt}{30\unitlength}\right\}$ \small block 2,,
,$\left.\rule{0pt}{30\unitlength}\right\}$ \small block 3,,
,$\left.\rule{0pt}{30\unitlength}\right\}$ \small block 4,,\kern30\unitlength$\;\vdots\;$](13)
\kern10\unitlength}\vskip10\unitlength
\caption{Encoding circuit for a rate $1/3$ convolutional quantum
  codes. Every gate has to be repeatedly applied shifted by one block,
  i.e. three positions down.  Note that the $\Csign$ gates are
  diagonal and hence can be arranged in any order.  For the $\Cnot$
  gates, each gate has to be repeated in its shifted version before the
  next gate can be applied.
\label{fig:example}}
\end{figure*}

\begin{corollary}
  Let $S(D)$ be the stabilizer matrix of a quantum convolutional code
  ${\cal C}$.  Then there exists a convolutional subcode ${\cal
    C}_{\text{sub}}\subseteq {\cal C}$ with a non-catastrophic encoder
  and encoder
  inverse such that asymptotically the rates of ${\cal
    C}_{\text{sub}}$ and ${\cal
  C}$ are equal.  Moreover, the encoder and its inverse only use 
Clifford gates and allow for online encoding and inverse encoding.
\end{corollary}

\section{Example}
Consider the $\F_4$-linear rate-$1/3$ convolutional code from
\cite[Table~VI]{FGG05}) with generator matrix
\[
G(D)=\left(
\begin{array}{ccc}
1+D & 1+\w D & 1+\wbar D 
\end{array}
\right).
\]
The corresponding stabilizer matrix is
\[
S(D)=\left(
\begin{array}{ccc|ccc}
  1+D&   1& 1+D&  0  &  D  & D\\
    0&   D&   D& 1+D & 1+D & 1
\end{array}
\right).
\]
The first sequence of $\overline{\Cnot}$ operations transforms the
first row of $X(D)$, and we obtain:
\[
\left(
\begin{array}{*{3}{c}|*{3}{c}}
1 & 0 & 0 & 1 & D & D \\
D^2 & D^2 + D & D^3 + D^2 + D & 0 & (D^2 + 1)/D & 1 
\end{array}
\right)
\]
Invertible row-operations do not change the stabilizer group, so
adding $D^2$ times the first row to the second yields
\begin{footnotesize}
\[\arraycolsep0.8\arraycolsep
\left(
\begin{array}{*{3}{c}|*{3}{c}}
1 & 0 & 0 & 1 & D & D \\
0 & D^2 + D & D^3 + D^2 + D & D^2 & (D^4 + D^2 + 1)/D & D^3 + 1 
\end{array}
\right).
\]
\end{footnotesize}%
Again using $\overline{\Cnot}$, we transform the second row of $X(D)$:
\begin{equation}\label{eq:neg_exponent}
\left(
\begin{array}{*{3}{c}|*{3}{c}}
1 & 0 & 0 & 1 & 1/D & (D^2 + D + 1)/D \\
0 & D & 0 & D^2 & 0 & D^3 + D^2 + D 
\end{array}
\right).
\end{equation}
Using $\overline{\Csign}$, we can clear the off-diagonal terms in the first row
of $Z(D)$,
\[
\left(
\begin{array}{*{3}{c}|*{3}{c}}
1 & 0 & 0 & 1 & 0 & 0 \\
0 & D & 0 & 0 & 0 & D^3 + D^2 + D 
\end{array}
\right),
\]
and similar for the second row
\[
\left(
\begin{array}{*{3}{c}|*{3}{c}}
1 & 0 & 0 & 1 & 0 & 0 \\
0 & D & 0 & 0 & 0 & 0 
\end{array}
\right).
\]
Finally, using $\overline{P}$ and $\overline{H}$ we get $Z$-only generators:
\[
\left(
\begin{array}{*{3}{c}|*{3}{c}}
0 & 0 & 0 & 1 & 0 & 0 \\
0 & 0 & 0 & 0 & D & 0 
\end{array}
\right)
\]
To clear the entries in the first row of the $Z$-part of
(\ref{eq:neg_exponent}), we need $\overline{\Csign}$ gates whose
target lie in the block before that of the target (see the third set
of gates in Fig.~\ref{fig:example}).  Therefore, encoding can only
start in the second block.  In the first block, all qubits are
initialized to $\ket{0}$.  Additionally, we ignore that the term $D$
in the second row implies that there is no operator $Z_i$ in the
stabilizer acting on the second qubit of the now second block and
constrain the input to $\ket{0}$.

A circuit for encoding is obtained by reversing the order of the
transformations.  The encoding circuit is illustrated in
Fig.~\ref{fig:example}.  Note that the circuit extends over three
blocks, i.e., has total memory two.  This is reflected by the fact
that in this example the operations used by the algorithm to clear
entries only involved Laurent polynomials of degree at most two.  In
contrast, an encoder for the classical convolutional code over $\F_4$
given by $G(D)$ can be realized with total memory one.

\magma{

g:=Matrix([[1+D,1+w*D,1+w^2*D]]);
G:=ChangeRing(XZmatrix(Matrix([g[1],w*g[1]])),F2);

G1:=G
*CNOT(2,4,3:R:=F2)
*CNOT(1,2,3:R:=F2)
*CNOT(1,3,3:R:=F2)
*CNOT(1,6,3:R:=F2);

G2:=AddRow(G1,D^2,1,2);
//G2:=G1;

G3:=G2
*CNOT(2,6,3:R:=F2)
*CNOT(3,5,3:R:=F2)
*CNOT(2,3,3:R:=F2);

G4:=MultiplyRow(G3,D,1);
//G4:=G3;

G5:=G4
*C_Z(4,2,3:R:=F2)
*C_Z(4,3,3:R:=F2)
*C_Z(4,6,3:R:=F2)
*C_Z(4,9,3:R:=F2);

G6:=G5
*C_Z(2,3,3:R:=F2)
*C_Z(2,6,3:R:=F2)
*C_Z(2,9,3:R:=F2);

G7:=G6
*P_Matrix(1,3:R:=F2)
*Hadamard(1,3:R:=F2)
*Hadamard(2,3:R:=F2);

}

\section{Conclusion}
We have shown that the quantum convolutional codes obtained from
self-orthogonal classical convolutional codes always have a subcode
which has asymptotically the same rate and allows for non-catastrophic
encoders.  This shows that errors affecting these codes do not
propagate in an unbounded fashion during the decoding process.  For
simplicity, we have presented the algorithm for qubit systems only,
but as in \cite{GRB:2003}, the technique applies to non-qubit
systems as well.


\section*{Acknowledgment}
The authors would like to thank Harold Ollivier and David Poulin for
fruitful comments on previous versions of the paper.



\IEEEtriggeratref{5}
\IEEEtriggercmd{\enlargethispage{-19.4cm}}


%



\end{document}